\documentclass[letterpaper,superscriptaddress,english,aps,prl,twocolumn,floatfix, showpacs, amsfonts, amssymb]{revtex4}
\usepackage{epsfig, graphicx,psfrag,amsmath,amssymb,float}
\usepackage[T1]{fontenc}
\usepackage[latin9]{inputenc}
\usepackage{amsmath}
\usepackage{amssymb}

\usepackage{amscd}
\usepackage{bm}
\usepackage{psfrag}

\usepackage{babel}

\newcommand{\be}{\begin{equation}}
\newcommand{\ee}{\end{equation}}
\newcommand{\bea}{\begin{eqnarray}}
\newcommand{\eea}{\end{eqnarray}}

\newcommand{\mc}{\mathcal}

\def\XXZ{\textit{XXZ}}

\begin{document}

\title{Low temperature dynamics  of nonlinear Luttinger liquids}
\author{C. Karrasch}
\affiliation{Department of Physics, University of California, Berkeley, CA 95720, USA }
\affiliation{Materials Sciences Division, Lawrence Berkeley National Laboratory, Berkeley, CA 94720, USA}
\author{R. G. Pereira}
\affiliation{Instituto de F\'{i}sica de S\~ao Carlos, Universidade de S\~ao Paulo, C.P. 369, S\~ao Carlos, SP,  13560-970, Brazil}
\author{J. Sirker}
\affiliation{Department of Physics and Research Center OPTIMAS, Technical University Kaiserslautern, D-67663 Kaiserslautern, Germany}
\affiliation{Department of Physics and Astronomy, University of Manitoba, Winnipeg, Canada R3T 2N2}
\date{\today}
\begin{abstract}
  We generalize nonlinear Luttinger liquid theory to describe the
  dynamics of one-dimensional quantum critical systems at low
  temperatures. Analyzing density-matrix renormalization group results
  for the spin autocorrelation function in the \XXZ\ chain we provide,
  in particular, direct evidence for spin diffusion in sharp contrast
  to the exponential decay in time predicted by conventional Luttinger
  liquid theory. Furthermore, we discuss how the frequencies and
  exponents of the oscillatory contributions from the band edges are
  renormalized by irrelevant interactions and obtain excellent
  agreement between our finite temperature nonlinear Luttinger liquid
  theory and the numerical data.
\end{abstract}


\maketitle

{\it Introduction}---The dynamics of quantum critical systems at
finite temperatures is an outstanding  challenge in
many-body physics \cite{sachdev}. Understanding the combined effects
of thermal fluctuations and interactions 
is crucial for the interpretation of experiments that probe
frequency-dependent responses and scattering cross sections.
Addressing this problem has become even more pressing since  fascinating 
new experiments in cold atomic gases
\cite{fukuhara,hild,knap,langen} and condensed matter
\cite{bisogni} have given us access to correlation functions directly
in the time domain.

Much of the interest in real time evolution of many-body states has
focused on one-dimensional (1D) models.  In particular, the difference
in the nonequilibrium dynamics of integrable versus nonintegrable
systems is currently a hotly debated topic
\cite{kinoshita,rigol,cazalilla,gangardt,fagotti}. In parallel,
studies of ground state dynamical correlations have recently forced us
to revise our understanding of quasiparticles in critical 1D systems
\cite{rozhkov,carmelo,pustilnik,pereira2006,khodas2007,imambekovscience,furusaki,matveev},
culminating with the development of the nonlinear Luttinger liquid
(NLL) theory \cite{imambekov}.  This theory provides a framework for
calculating exponents of edge singularities in dynamical response
functions based on a picture of fermionic quasiparticles with
nonlinear dispersion, reminiscent of elementary excitations in Bethe
ansatz (BA) solvable models \cite{pereira2008,essler10}. NLL theory
predicts that the long-time decay of correlation functions is
dominated by excitations involving particles or holes near band edges,
explaining the high-frequency oscillations observed numerically
\cite{pereira2008,pereira2009} and confirmed by an exact form factor
approach \cite{kitanine}.

The purpose of this Letter is to extend NLL theory to describe the low
temperature, long time decay of correlation functions of 1D quantum
fluids. Determining the precise effects of temperature is certainly
relevant for experiments aimed at testing the phenomenology of NLLs.
Another motivation for this work is to provide analytic expressions to
examine numerical results for time dependent correlation functions,
accessible by finite-temperature versions \cite{sirkerklumper,barthel,karrasch,karrasch2,barthel2} of time-dependent \cite{tdmrg1,tdmrg2,tdmrg3,tdmrg4,tdmrg5} density
matrix renormalization group (tDMRG) methods \cite{white,dmrgrev}, and for the thermal
broadening of edge singularities in the frequency domain
\cite{panfil}. In the following we will generalize NLL theory to $T>0$
and compare the predictions with state-of-the-art tDMRG
results. 
We will concentrate on the spin autocorrelation function $G(t)$ of the
\XXZ\ model at zero magnetic field, but our approach can easily be
generalized to other correlations and 1D models. We stress that the
time decay of correlation functions in the \XXZ\ model at finite $T$
is still an open problem despite the integrability of the model
\cite{korepin,korepinbook}. Our main results are: ({\it i}) $G(t)$ at
intermediate times $t$ is well described by a generalization of NLL
theory that takes into account the effects of irrelevant operators on
the dispersion of high-energy quasiparticles and on their coupling to
low energy modes; ({\it ii}) at long times, $G(t)$ contains a
non-oscillating, $\sim 1/\sqrt t$ decaying diffusive term, in sharp
contrast to the exponential decay predicted by Luttinger liquid
theory.  Diffusive behavior in spins chains has been invoked to
explain the spin-lattice relaxation rate \cite{thurber} and muon-spin
relaxation \cite{xiao} measured in quasi-1D antiferromagnets and is
attributed to inelastic umklapp scattering \cite{sirker}. However,
fingerprints of spin diffusion have so far only been found indirectly
in the current-current correlation function \cite{sirker,karrasch,drudepaper2} and
in the imaginary time dependence of the dynamical susceptibility
\cite{brenig}. Here we provide the first direct numerical evidence of
spin diffusion in the autocorrelation function.

{\it Model}---Consider the 1D \XXZ\ Hamiltonian \be
H=\sum_{j=1}^L\left(S_j^xS_{j+1}^x+S_j^yS_{j+1}^y+\Delta
  S_j^zS_{j+1}^z\right),\label{xxz} \ee where $S_j^a$ are spin-$1/2$
operators, $\Delta$ is the anisotropy parameter, and periodic boundary
conditions are assumed. This model can be realized, for instance, in
the Mott insulating phase of two-component Bose mixtures in 1D optical
lattices \cite{duan,fukuhara}. The spin autocorrelation function at
temperature $T$ is defined by\be G(t)=\langle
S^z_j(t)S^z_j(0)\rangle=\textrm{Tr}\{ S^z_j(t)S^z_j(0)e^{-H/T}\}/Z,
\ee where $Z=\textrm{Tr}\,e^{-H/T}$ is the partition function. The
\XXZ\ model is equivalent to spinless fermions, $c^{(\dagger)}_j$, via
a Jordan-Wigner transformation with $S_j^z\to c_j^\dagger
c^{\phantom\dagger}_j-1/2$, so that $\Delta$ plays the role of a
nearest-neighbor density-density interaction \cite{giamarchi}.

{\it Noninteracting case}---For $\Delta=0$, the \XXZ\ model reduces
to a free fermion model. The autocorrelation factorizes
into a product of free particle and hole Green's functions and is
given exactly by \cite{sirkerprb2006} \be G^{(0)}(t)=\langle
c_j^{\phantom\dagger}(t)c_j^{\dagger}(0)\rangle\langle
c_j^{\dagger}(t)c_j^{\phantom\dagger}(0)\rangle=\left(\int_{-\pi}^{\pi}\frac{dk}{2\pi}\,
  f_ke^{i\epsilon_kt}\right)^2 \label{G0}, \ee with dispersion
$\epsilon_k=-\cos k$ and Fermi distribution
$f_k=1/(e^{\epsilon_k/T}+1)$ at half-filling. 

Let us discuss an approximation that captures the asymptotic long time
decay of $G^{(0)}(t)$. First note that $G^{(0)}(t)$ oscillates at
arbitrary temperatures due to saddle points of the integrand where
$d\epsilon_k/dk=0$, which occur at $k=0$ and $k=\pi$.  In the
low-temperature regime $T\ll 1$, we employ a mode expansion of the
fermion field which keeps only states within sub-bands near the
smeared Fermi surface and near the saddle points, in the form
$c_{j=x}\sim \psi_R(x)e^{i\pi x/2}+ \psi_L(x)e^{-i\pi x/2}+ \bar
d^\dagger(x)+e^{i\pi x}d(x)$. Here $\psi_{R,L}$ are the low-energy
right- and left-moving components, while $\bar d$ and $ d$ are
high-energy modes: $\bar d^\dagger $ creates a hole at the bottom of
the band ($k=0$), and $ d$ annihilates a particle at the top of the
band ($k=\pi$).

The mode expansion reduces the problem to the calculation of free
propagators for $\psi_{R,L}$ and $d,\bar d$. The low energy modes have
linear dispersion, thus their propagator at $T\ll 1$ is given by the
standard conformal field theory result $\langle
\psi^{\phantom\dagger}_{R,L}(x,t)\psi^{\dagger}_{R,L}(x,0) \rangle
\sim \pi T/\sinh(\pi T t)$. On the other hand, the high-energy modes
do not feel the temperature, up to an exponentially small correction
in the Fermi distribution. Since the dispersion is parabolic near the
band edges, $\epsilon_k\approx 1-k^2/2$, the propagator of $d,\bar d$
decays slowly, $\langle d (x,t) d^\dagger(x,0)\sim e^{-it}/\sqrt{t}$.
Substituting the contributions from low and high energy modes into Eq.
(\ref{G0}), we obtain the asymptotic decay of $G^{(0)}(t)$: \be
G^{(0)}(t)\approx \frac{i e^{-i2t}}{2\pi
  t}+\frac{\sqrt{2}Te^{-i(t+\pi/4)}}{\sqrt{\pi t}\sinh (\pi T
  t)}-\frac{T^2}{\sinh^2(\pi T t)}.\label{G0finiteT} \ee The first
term stems from the contribution in which both particle and hole are
high-energy modes. The second term is due to a high-energy particle
(hole) plus a low-energy hole (particle) at either one of the Fermi
points.  The latter decays exponentially for $t\gg 1/T$. The third
contribution is due only to the low-energy modes $\psi_{R,L}$, and
decays more rapidly than the oscillating terms. Therefore,
$G^{(0)}(t)$ for $t\gg 1/T$ is dominated by the term oscillating with
frequency $\omega=2$ and decaying as $1/t$.
   
{\it Interacting case: NLL theory}---We focus on the regime $0<\Delta<1$
corresponding to a critical phase of fermions with repulsive
interactions.  The factorization of $G(t)$ as in Eq.~(\ref{G0}) is then \emph{a priori} lost.
However, we can investigate the long time decay of $G(t)$  using the framework of NLL theory  \cite{imambekov}. The idea is to  keep the mode
expansion with both low and high energy modes, which are now
identified as quasiparticles in a renormalized band. The ground state
is a vacuum of $d$ and $\bar d$, and the spin operator $S_j^z\sim c^\dagger_j c_j^{\phantom\dagger}$ creates
at most one $d$ particle and/or one $\bar d$ hole. At low temperatures, we  neglect an exponentially small thermal population of the band edge modes. The $d,\bar d$
quasiparticles can then be treated as mobile ``impurities''
distinguishable from the low energy modes. 

At $T=0$ the dynamics is described by an effective field theory with
Hamiltonian density \cite{pereira2008} \bea \mc H&=&d^\dagger
\left(\varepsilon+\frac{\partial_x^2}{2m}\right)d +\bar d^\dagger
\left(\varepsilon+\frac{\partial_x^2}{2m}\right)\bar d+Vd^\dagger
d\,\bar d^\dagger \bar d
\nonumber\\
\!\!\!\!&+&\frac v2[(\partial_x\theta)^2+(\partial_x\phi)^2]+
\frac{v\alpha}{\sqrt{\pi K}}\partial_x\phi (d^\dagger d-\bar d^\dagger
\bar d).\label{impuritymodel} \eea Here $\varepsilon $ is the energy
and $m$ the absolute value of the effective mass of the band edge
modes, $V$ is the impurity-impurity interaction, $\phi(x)$ and
$\theta(x)$ are dual fields representing the bosonized low-energy
modes \cite{giamarchi} and obey
$[\phi(x),\partial_{x^\prime}\theta(x^\prime)]=i\delta(x-x^\prime)$.
$v$ is the spin velocity, $K$ the Luttinger parameter, and $\alpha$
the dimensionless coupling constant of the impurity-boson interaction.
For the integrable \XXZ\ model one finds
$v=\varepsilon=\frac1m=\frac{\pi\sqrt{1-\Delta^2}}{2\arccos\Delta}$,
$K=1-\frac{\alpha}{2\pi}=\frac{\pi/2}{\pi-\arccos\Delta}$, while
$V\approx -4\Delta$ for $\Delta\ll 1$.

Eq. (\ref{impuritymodel}) must be regarded as a fixed point Hamiltonian which includes all \emph{marginal} interactions
allowed by symmetry. This model can be solved exactly by performing a
unitary transformation that decouples the impurities from the bosonic
modes, but attaches ``string'' operators to the $d,\bar d$ fields, {\it i.e.} $d(x)\to d(x)e^{-i\alpha \theta(x) /\sqrt{\pi K}}$
\cite{pustilnik}. This allows one to predict the   long time decay of $G(t)$ at $T=0$, up to non-universal amplitudes \cite{pereira2008}. 

We now extend NLL theory to $0<T\ll \varepsilon$ (low temperatures
compared to the renormalized bandwidth), using
Eq.~(\ref{impuritymodel}) as a starting point. We obtain the leading
$T$ dependence by analyzing the effects of \emph{irrelevant}
interactions. The leading corrections to the oscillating terms in
$G(t)$ allowed by symmetry \cite{supplem} stem from the irrelevant
dimension-three operators \bea
\delta \mc H&=& g[(\partial_x\theta)^2+(\partial_x\phi)^2](d^\dagger d+\bar d^\dagger \bar d)\nonumber\\
&&+g^\prime[(\partial_x\theta)^2-(\partial_x\phi)^2](d^\dagger d+\bar
d^\dagger \bar d)\nonumber\\
&&-\mu_{+}\partial^2_{x}\theta\left(d^{\dagger}d-\bar d^{\dagger}\bar d\right)\nonumber \\
&&+\mu_{-}\partial_{x}\theta\left(-id^{\dagger}\partial_{x}d+i\bar
  d^{\dagger}\partial_{x}\bar d+h.c.\right). \label{irrelOp1} \eea
Substituting the mode expansion into Hamiltonian (\ref{xxz}) and
bosonizing the low-energy modes, we find for $\Delta\ll 1$: $g\approx
-\Delta$ , $\mu_-\approx- \Delta/\sqrt{\pi}$, while $g^\prime$ and
$\mu_+$ are not generated to first order in $\Delta$. The $g^\prime$
interaction is particularly important: it can be identified with a
three-body scattering process \cite{khodas2007,pereira2008} and gives
rise to a nonzero impurity decay rate for $T>0$
\cite{castroneto,karzig}. However, by imposing nontrivial conservation
laws in the \XXZ\ model we can show that $g^\prime=0$ exactly
\cite{supplem,KluemperSakai}, as expected from the lack of three-body
scattering in integrable models.

We calculate the impurity self-energy $\Sigma$ and effective
impurity-boson interaction $\tilde \alpha$ by perturbation theory in
the irrelevant operators. The leading $T$ dependence is determined by
loop diagrams which contain only one irrelevant coupling constant but
arbitrary factors of the bare $\alpha$. In the calculation of loop
diagrams, it is convenient to treat the quadratic term in the impurity
dispersion (which is also a dimension-three operator) as a
perturbation, expanding the internal impurity propagators in powers of
$1/m$. We find\be \Sigma(T)\approx c_\Sigma T^2/v^2, \quad
\tilde\alpha(T)\approx \alpha (1+c_\alpha T/v^2).\label{Tcorrections}
\ee The prefactors $c_\Sigma$ and $c_\alpha$ are linear functions of
$g,\mu_+,\mu_-$. Bearing in mind that $g,\mu_-\sim \mc O(\Delta)$,
$\mu_+\sim \mc O(\Delta^2)$ for $\Delta\ll 1$ , we obtain the weak
coupling approximation\be c_\Sigma\approx \frac{\pi g}3
+\frac{\alpha^2}{12 Km},\quad c_\alpha\approx -2g+\frac{\alpha^2}{2\pi
  Km}.\label{prefactors} \ee Omitted terms are $\mc O(\Delta^3)$ or
higher. To first order in $\Delta$, the self-energy and vertex
correction are both governed by $g\approx -\Delta$. For $0<\Delta\ll
1$, Eq. (\ref{Tcorrections}) thus implies that the effective impurity
energy $\tilde \varepsilon(T)=\varepsilon+\Sigma(T)$ decreases $\sim
T^2$, whereas the effective coupling $\tilde \alpha$ increases
linearly with $T$. Phenomenologically, we find $g=\frac{\pi
  v^2}{2K}\frac{\partial^2\varepsilon}{\partial h^2}\big|_{h=0}$
relating the coupling constant $g$ in Eq.~(\ref{irrelOp1}) to a change
in the {impurity energy} when applying a magnetic field $h$. This
allows one to obtain $g(\Delta)$ exactly using the BA and Wiener-Hopf
techniques. Unfortunately, the corrections of higher order in $\alpha$ in
Eq.~(\ref{Tcorrections}) quickly become of the same order as the
$\mathcal{O}(g)$ term making it impossible to fix $c_\Sigma$,
$c_\alpha$ beyond the lowest order in $\Delta$.  Importantly however,
the $T$ dependence does hold for arbitrary $0<\Delta<1$, as long as
the temperature is small enough.

The renormalization of the impurity-impurity interaction appears at
order $g^2$. The basic process involves a two-boson loop connecting
two impurity lines. The correction depends on the momentum and
frequency exchange between the impurities: $\tilde
V(q,\omega,T)\approx V+\frac{2\pi g^2T^2\omega}{3v^4q}$, where we
simplified the result in the physically relevant regime $\omega\ll vq$
for impurities with parabolic dispersion.

We can now describe the decay of $G(t)$ using the methods of NLL
theory with renormalized parameters at $T>0$. First, consider the
contribution from the excitation with a single impurity, equivalent to
the second term in Eq. (\ref{G0finiteT}). The unitary transformation
introduces a ``string'' operator whose scaling dimension depends on
temperature through $\tilde \alpha(T)$. The result is \be
G_{1}(t)\approx \frac{ A(T)}{\sqrt{t}}\left[\frac{\pi T}{\sinh(\pi T
    t)}\right]^{\tilde\eta(T)}e^{-i[\tilde\varepsilon(T)t+\tilde\varphi(T)]},
\label{GtT}
\ee where
$\tilde\eta(T)=\frac{K}2+\frac1{2K}\left[1-\frac{\tilde\alpha(T)}{2\pi}\right]^2$,
$\tilde\varphi(T)=\frac{\pi}{2}\left[\tilde\eta(T)-\frac12\right]$,
and $  A(T)$ is the unknown prefactor. Note that $\tilde
\alpha(T)>\alpha$ for $0<\Delta\ll1$ implies that  $\tilde\eta(T)$ decreases with temperature,  slightly slowing down the   decay of  $G_{1}(t)$.

The two-impurity contribution to $G(t)$, analogous to the first term
in Eq. (\ref{G0finiteT}), is strongly modified by the $V$ interaction.
At $T=0$, the $1/t$ decay for $\Delta =0$ changes to $1/t^2$ for
$\Delta >0$ and $t\gg 1/V^2$ \cite{pereira2008}. This asymptotic
behavior is associated with two impurities scattering in a ladder
series with small energy and momentum transfer, $\omega\sim q^2/m\ll
\varepsilon$.  Neglecting the renormalization of $V$ for $\omega\ll
vq$, we obtain the two-impurity contribution \be G_{2}(t)\approx
\frac{B(T)}{t^2}\, e^{-i2\tilde \varepsilon(T)t},\label{g2imp} \ee
where $B(T)$ is the unknown prefactor. For generic models,
Eq.~(\ref{g2imp}) must be modified to include the exponential decay
due to relaxation by the three-body process $g^\prime$.

{\it Diffusive decay}---At temperatures $T\ll 1$, conventional
Luttinger liquid theory predicts that the non-oscillating terms in
$G(t)$, associated only with the low-energy modes $\psi_{R,L}$, are
given in the interacting case by $ A^\prime[\pi T/\sinh(\pi T
t)]^2+B^\prime[\pi T/\sinh(\pi T t)]^{2K}$, where the amplitudes
$A^\prime,B^\prime$ are known \cite{lukyanov}. However, we must also
consider how irrelevant operators affect the low-energy contributions.
In \cite{sirker} it was shown that the formally irrelevant umklapp
scattering \be \delta\mathcal{H}_U =\lambda \cos(4\sqrt{\pi K}\phi)
\label{Umklapp}
\ee qualitatively changes the long time decay of the low-energy,
long-wavelength contribution to $G(t)$.  There appears a new time
scale set by the decay rate $\gamma\propto \lambda^2T^{8K-3}$. For the
\XXZ\ model, $\gamma$ can be calculated exactly
\cite{lukyanov,sirker}. At low $T$ and long times $t\gg 1/\gamma\gg
1/T$, we find a diffusive contribution \footnote{Note that this
  corrects the result in \cite{sirker} where a factor $2\pi$ is
  missing.}  \be G_{\textrm{diff}}(t)=\frac{\Gamma}{\sqrt{t}}\quad
,\quad \Gamma=\frac{KT}{\pi v^2}\sqrt{\frac{\gamma}{2\pi}}.
\label{Gdiff}\ee
Strikingly, this diffusive ($\sim 1/\sqrt{t}$) decay in the
autocorrelation function coexists with ballistic transport
\cite{sirker,prosen}. Eqs.~(\ref{GtT}, \ref{Gdiff}) are the two
contributions to $G(t)$ which are dominant at intermediate and long
times, respectively.

{\it Numerical results}---In order to test our theory, we now turn to
a comparison with tDMRG results for finite system size.  Non-zero
temperatures are incorporated via a purification of the density
matrix. By using the disentangler introduced in Ref.~\cite{karrasch}
and exploiting time translation invariance $\langle
S^z_j(t)S^z_j(0)\rangle=\langle S^z_j(t/2)S^z_j(-t/2)\rangle$
\cite{barthel2} we can substantially extend the accessible time scale.
Details of the algorithm are described in
Refs.~\cite{karrasch2,dmrgrev}. By varying both the system size and
the bound for the maximally discarded weight we ensure that the finite
size and truncation errors are smaller than the symbol size for all
data presented in the following.
\begin{figure}
\begin{center}
\includegraphics*[width=.99\columnwidth]{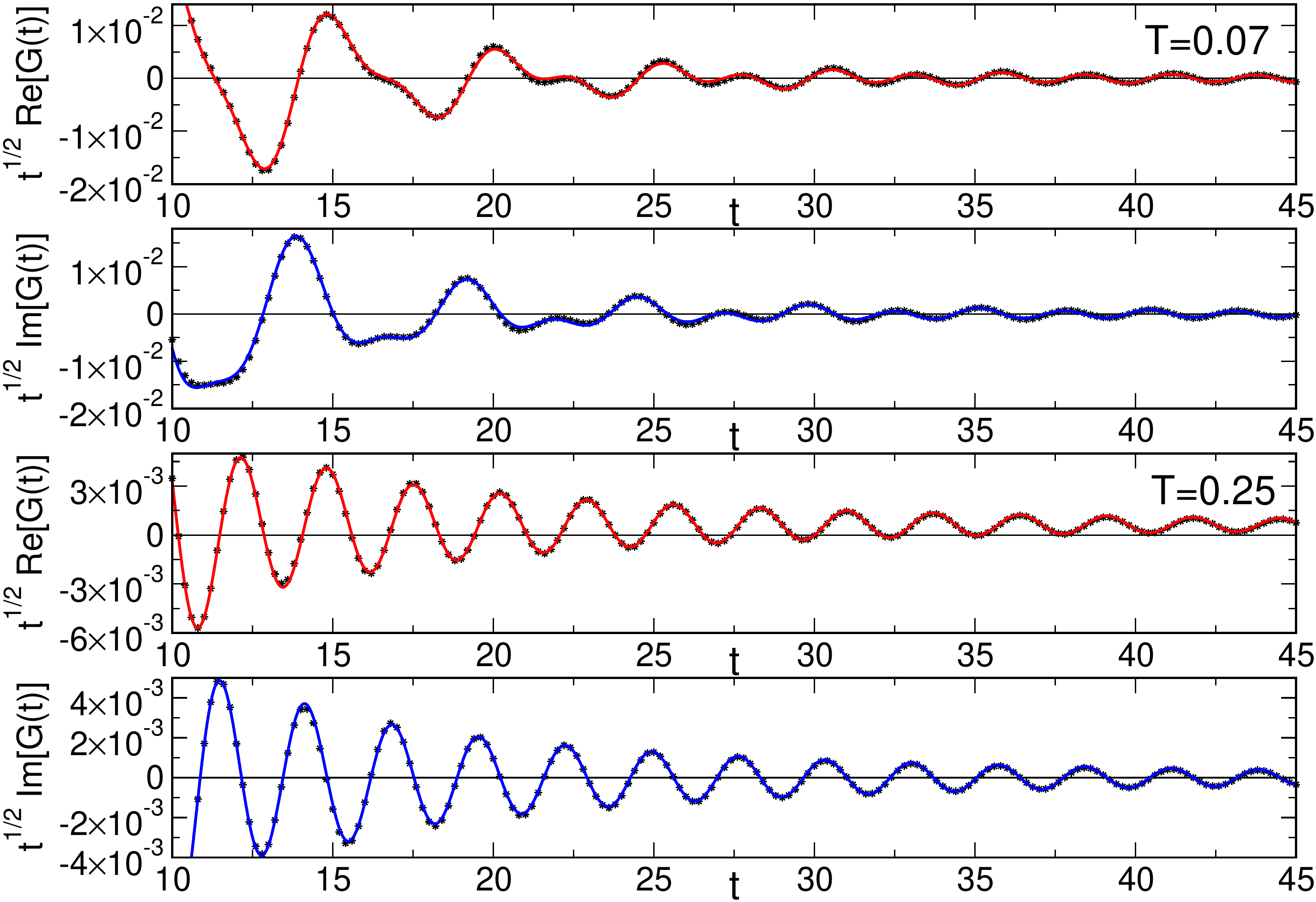}
\end{center}
\caption{$\sqrt{t}G(t)$ at $\Delta=0.3$: tDMRG data (symbols) and fits
  (lines) for $t\geq 10$. Here
  $\lim_{t\to\infty}\sqrt{t}G(t)=\Gamma$ with $\Gamma_\textrm
  {fit}(T=0.07)\approx 0$ [$\Gamma_\textrm{th}(T=0.07)=2.1\cdot
  10^{-5}$] and $\Gamma_\textrm {fit}(T=0.25)\approx 6.3\cdot 10^{-4}$
  [$\Gamma_\textrm{th}(T=0.25)=7.9\cdot 10^{-4}$].}
\label{delta05}
\end{figure}
\begin{figure}
\begin{center}
\includegraphics*[width=.99\columnwidth]{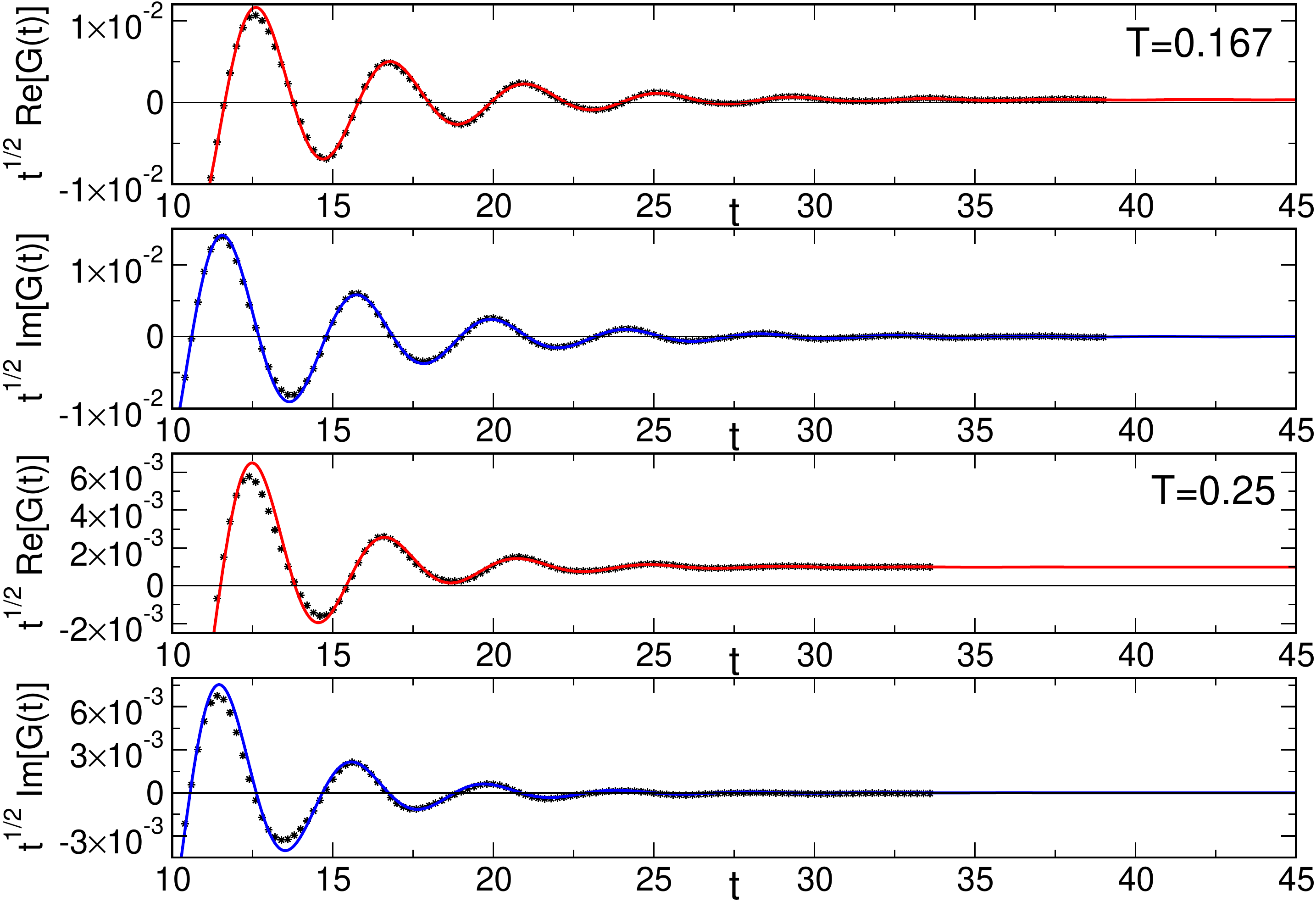}
\end{center}
\caption{Same as Fig.~\ref{delta05} for $\Delta=0.8$ with $\Gamma_\textrm {fit}(T=0.167)\approx 3.3\cdot 10^{-4}$
  [$\Gamma_\textrm{th}(T=0.167)=5.6\cdot 10^{-4}$] and $\Gamma_\textrm
  {fit}(T=0.25)\approx 9.9\cdot 10^{-4}$ [$\Gamma_\textrm{th}(T=0.25)=1.3\cdot
  10^{-3}$].}
\label{delta08}
\end{figure}

\begin{figure}
\begin{center}
\includegraphics*[width=.99\columnwidth]{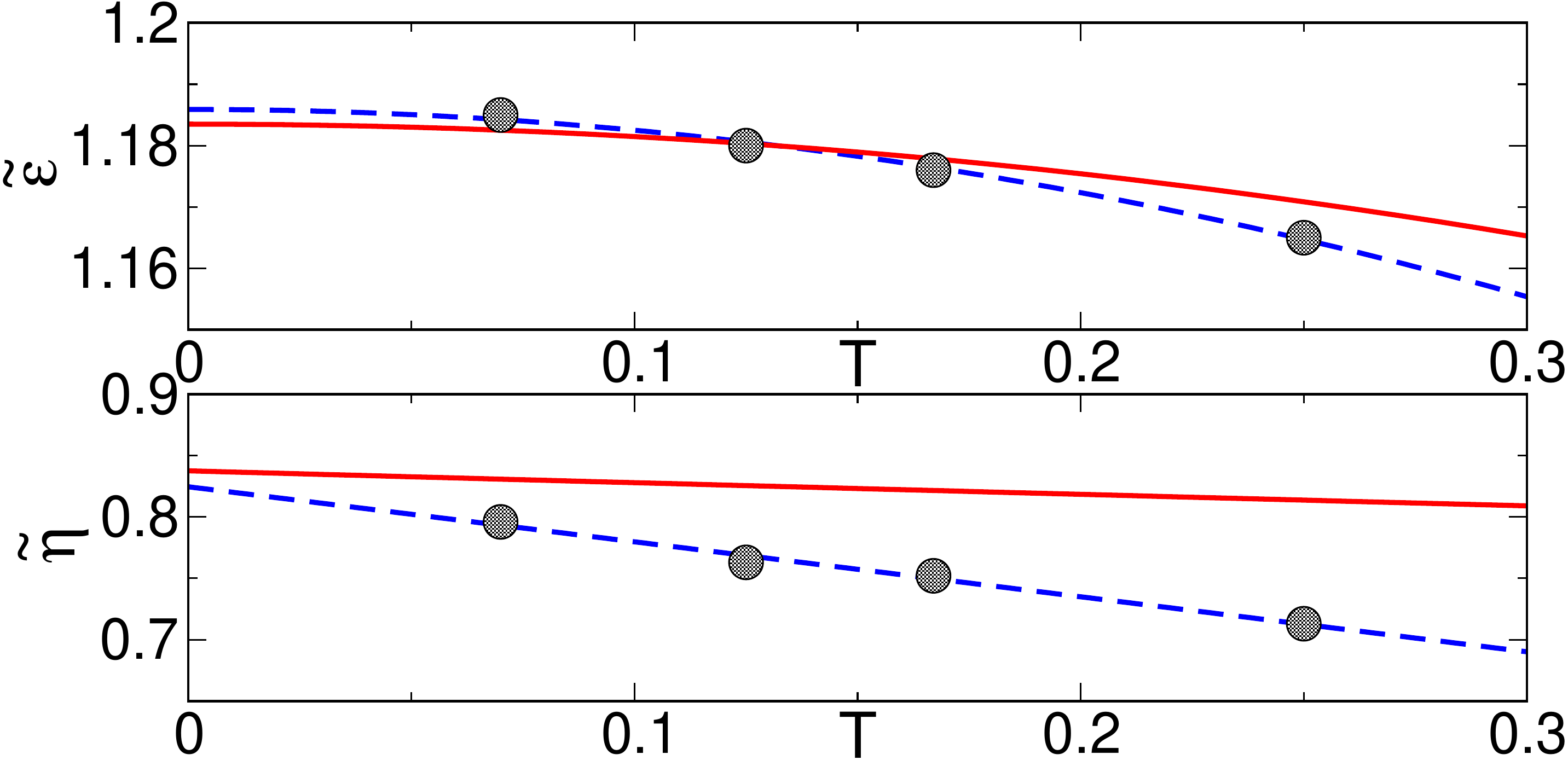}
\end{center}
\caption{$\tilde\varepsilon(T)$ and $\tilde\eta(T)$ for $\Delta=0.3$
  from fits of tDMRG data (symbols). Dashed lines: quadratic
  and linear least square fits, respectively; solid lines: 
  $\mathcal{O}(\Delta)$ theoretical result, Eq.~\eqref{Tcorrections}.}
\label{parameters}
\end{figure}

We find a striking difference between the weakly interacting case,
Fig.~\ref{delta05}, and the strongly interacting case,
Fig.~\ref{delta08}. While the data for $\Delta=0.8$ can be very well
fitted by a sum of the single impurity contribution \eqref{GtT} and
the diffusive part \eqref{Gdiff}, it is necessary to also include the
two-impurity contribution \eqref{g2imp} for $\Delta=0.3$. In the
latter case, we also allow for a decay rate $\rho$ by multiplying
Eq.~(\ref{g2imp}) by $e^{-\rho t}$. The fits yield very small decay
rates which seem to be of order $\sim\text{e}^{-1/T}$ and could
possibly be related to thermal excitations at the band edges which we
have neglected in our analysis \cite{supplem}. At $T=0.25$ we find for
both $\Delta$ values clear evidence for spin diffusion with diffusion
constants $\Gamma_{\textrm{fit}}$ close to the theoretically predicted
values, $\Gamma_{\textrm{th}}$, see Eq.~\eqref{Gdiff}. In agreement
with theory we also find numerically that the diffusive contribution
seems to vanish for magnetic fields $h\gg T$ (data not shown) where the Umklapp term
\eqref{Umklapp} is oscillating and should be dropped from the
effective theory  \cite{sirker}. The oscillation
frequency $\tilde\varepsilon(T)$ and exponent $\tilde\eta(T)$,
obtained from fits of tDMRG data at $\Delta=0.3$, are shown in
Fig.~\ref{parameters}.
The data confirm that $\tilde\varepsilon\sim -T^2$ and $\tilde\eta\sim
-T$ while the prefactors, even for $\Delta=0.3$, already seem to
deviate significantly from the lowest order result,
Eq.~\eqref{prefactors}.

{\it Exact parameters}---The integrability of the \XXZ\ model raises
the question whether parameters such as the frequency $\tilde
\varepsilon(T)$ can be determined exactly, beyond the lowest order in
Eq.~(\ref{Tcorrections}). At $T=0$ the exact $\varepsilon$ is
identified with the half bandwidth of the elementary excitations
computable by BA \cite{pereira2008}. The natural extension to $T>0$
should be based on the energies of excitations on top of an
equilibrium state in the thermodynamic Bethe ansatz (TBA) approach
\cite{takahashi72,puga}. However, by solving TBA nonlinear integral
equations numerically \cite{supplem} we find that the bandwidth of the
dressed energy for a single-hole excitation increases with $T$,
whereas the perturbative expression (\ref{Tcorrections}) and the tDMRG
results in Fig.~\ref{parameters} show that $\tilde\varepsilon(T)$
decreases with $T$. This disagreement is rather puzzling given that a
generalized TBA approach has been shown to be applicable even to
nonequilibrium dynamics \cite{caux}.

{\it Conclusion}---Finite temperatures and quantum fluctuations lead
to large non-perturbative effects on dynamical correlations in NLL's.
The exponents of oscillating contributions are renormalized by
temperature while umklapp scattering leads to spin diffusion
dominating the long-time asymptotics. These predictions are in
excellent agreement with tDMRG calculations which provide, in
particular, the first direct evidence for spin diffusion in the \XXZ\
model and show striking changes in the one- and two-impurity
contributions as a function of temperature and interaction strength.
The theory is easily extended to models such as the Bose gas and can
also be used to study the propagation of an impurity through a 1D
quantum fluid at finite temperatures. We expect, in particular, that
these results will help to set up and interpret experiments in
ultracold atoms aiming at an observation of spin diffusion
\cite{knap,hild}.

We thank F.H.L.~Essler, L.I. Glazman, A.~Kl\"umper and M.~Panfil for helpful
discussions. J.S.~acknowledges support by the Collaborative Research
Centre SFB/TR49, the Graduate School of Excellence MAINZ (DFG,
Germany), as well as NSERC (Canada). R.G.P. acknowledges support by CNPq (Brazil). C.K.~acknowledges support by Nanostructured Thermoelectrics program of LBNL.

\clearpage
\appendix

\onecolumngrid
\setcounter{equation}{0}
\begin{center}
\LARGE{Supplemental Material:\\ ``Low temperature dynamics of nonlinear Luttinger liquids''\\ by C.~Karrasch, R.~G.~Pereira, and
  J.~Sirker}
\end{center}

\subsection{1. Irrelevant impurity-boson interactions}

In this section we discuss how discrete symmetries and integrability of the XXZ model constrain the coupling constants of irrelevant operators in the effective impurity model.

Particle-hole ($C$) and parity ($P$) transformations act on the bosonic fields and impurity fields as follows \cite{lukyanov} :
\begin{equation}
C:\left\{ \begin{array}{ccc}
\theta&\to&-\theta\\
\phi&\to&-\phi\\
d&\to &\bar d\\
\bar d&\to &  d
\end{array}\right.\quad,\quad P:\left\{ \begin{array}{ccc}
x&\to&-x\\
\theta&\to&\theta\\
\phi&\to&-\phi\\
d&\to& d\\
\bar d&\to& \bar d\end{array}\right..
\end{equation}
The Hamiltonian  can  only contain  operators that are invariant under $C$ and $P$. Let us denote by $H_{a}^{(n)}=\int dx\, \mc H_{a}^{(n)}(x)$, $a=1,2,\dots$, a particular term in the Hamiltonian such that $\mc H_a^{(n)}$ is an operator with scaling dimension $n$. Up to dimension four, we have a list of 14 operators allowed by symmetry:
\bea
\mathcal{H}_{1}^{(2)} & =&\frac{v}{2}\left[(\partial_{x}\theta)^{2}+(\partial_{x}\phi)^{2}\right],\nonumber \\
\mathcal{H}_{2}^{(1)} & =&\varepsilon \left(d^{\dagger}d+\bar d^{\dagger}\bar d\right),\nonumber \\
\mathcal{H}_{3}^{(3)} & =&-\frac{1}{2m}\left(\partial_{x}d^{\dagger}\partial_{x}d+\partial_{x}\bar d^{\dagger}\partial_{x}\bar d\right),\nonumber \\
\mathcal{H}_{4}^{(2)} & =&\frac{ v\alpha}{\sqrt{\pi K}}\partial_{x}\phi\left(d^{\dagger}d-\bar d^{\dagger}\bar d\right),\nonumber \\
\mathcal{H}_{5}^{(2)} & =&-Vd^{\dagger}d\bar d^{\dagger}\bar d,\nonumber \\
\mathcal{H}_{6}^{(3)} & =&g\left[(\partial_{x}\theta)^{2}+(\partial_{x}\phi)^{2}\right]\left(d^{\dagger}d+\bar d^{\dagger}\bar d\right),\nonumber \\
\mathcal{H}_{7}^{(3)} & =&g^\prime\left[(\partial_{x}\theta)^{2}-(\partial_{x}\phi)^{2}\right]\left(d^{\dagger}d+\bar d^{\dagger}\bar d,\right) \label{eq:allHs} \\
\mathcal{H}_{8}^{(3)} & =&\mu_{+}\partial_{x}\theta\,\partial_{x}\left(d^{\dagger}d-\bar d^{\dagger}\bar d\right),\nonumber \\
\mathcal{H}_{9}^{(3)} & =&-i\mu_{-}\partial_{x}\theta\left(d^{\dagger}\partial_{x}d-\partial_{x}d^{\dagger}d-\bar d^{\dagger}\partial_{x}\bar d+\partial_{x}\bar d^{\dagger}\bar d\right),\nonumber \\
\mathcal{H}_{10}^{(4)} & =&\vartheta_{+}\left(d^{\dagger}\partial_{x}d\partial_{x}d^{\dagger}d+\bar d^{\dagger}\partial_{x}\bar d\partial_{x}\bar d^{\dagger}\bar d\right),\nonumber \\
\mathcal{H}_{11}^{(4)} & =&\vartheta_{-}\left(d^{\dagger}\partial_{x}d-\partial_{x}d^{\dagger}d\right)\left(\bar d^{\dagger}\partial_{x}\bar d-\partial_{x}\bar d^{\dagger}\bar d\right),\nonumber \\
\mathcal{H}_{12}^{(4)} & =&i\kappa_{1}\partial_{x}^{2}\phi\left(d^{\dagger}\partial_{x}d-\partial_{x}d^{\dagger}d-\bar d^{\dagger}\partial_{x}\bar d+\partial_{x}\bar d^{\dagger}\bar d\right),\nonumber \\
\mathcal{H}_{13}^{(4)} & =&i\kappa_{2}\partial_{x}\phi\partial_{x}\theta\left(d^{\dagger}\partial_{x}d-\partial_{x}d^{\dagger}d+\bar d^{\dagger}\partial_{x}\bar d-\partial_{x}\bar d^{\dagger}\bar d\right),\nonumber \\
\mathcal{H}_{14}^{(4)} & =&\kappa_{3}\partial_{x}\phi\left(\partial_{x}d^{\dagger}\partial_{x}d-\partial_{x}\bar d^{\dagger}\partial_{x}\bar d\right).\nonumber
\eea
All the operators in Eq. (\ref{eq:allHs})  conserve  the total number of particles and holes in the high-energy subbands.  In the above list we have omitted  operators which couple the impurity modes to umklapp type operators. The lowest dimension operator in this family is  $( d^\dagger d+\bar d^\dagger \bar d)\cos(4{\sqrt{\pi K}\phi})$, whose  scaling dimension varies continuously  from $5$ at  $\Delta=0$ to $3$ at $\Delta=1$. The set of constraints on the coupling constants that   we shall derive in the following is not affected by this family of operators.

The integrability of the XXZ model affects the effective impurity model by constraining the coupling constants of irrelevant interactions. Following \cite{pereira2006},   we shall examine the consequences of integrability by imposing the existence of nontrivial conservation laws. For the XXZ model, it is fortunate that the first nontrivial conserved quantity can be identified with the energy current operator $J_E$, which is defined from the continuity equation for the Hamiltonian density \cite{KluemperSakai}. In the continuum limit, the energy current density is given by \be
\partial_xj(x)=i[\mc H(x),H],
\ee
where $H=\int dx\, \mc H(x)=\sum_{a}\int dx\, \mc H_a$ is the total Hamiltonian. The energy current operator is $J_E=\int dx\, j(x)$. Let us denote by $j_{a,b}$, with $a<b$, the contribution to the energy current density  obtained by taking the commutator of   terms $\mc H_a$ and $\mc H_b$ in the Hamiltonian as follows: 
\begin{equation}
\partial_{x}j_{a,b}^{(n+m-2)}=i\int dy\,\left\{ \left[\mathcal{H}_{a}^{(n)}(x)\mbox{,}\mathcal{H}_{b}^{(m)}(y)\right]+\left[\mathcal{H}_{b}^{(m)}(x)\mbox{,}\mathcal{H}_{a}^{(n)}(y)\right]-\delta_{ab}\left[\mathcal{H}_{a}^{(n)}(x)\mbox{,}\mathcal{H}_{a}^{(n)}(y)\right]\right\}.
\end{equation}
The notation  implies that when we take the commutator of a dimension-$n$ operator
with another dimension-$m$ operator, the corresponding contribution to
$J_{E}$ (when nonvanishing) has dimension $n+m-2$. 

To check that $J_{E}=\sum_{a<b}J_{a,b}$ is conserved, we need to take the commutator
with all the terms in $H$ again:
\begin{equation}
\left[J_{a,b}^{(n+m-2)},H_{c}^{(l)}\right]=\int dx\,\mathcal{O}_{a,b,c}^{(n+m+l-1)}(x).\label{commJH}
\end{equation}
We organize the expansion by operator dimension. In order to find nontrivial relations between the coupling constants in Eq. (\ref{eq:allHs}), it suffices to compute $[J_E,H]$ to the level of dimension-four operators.  For this we  need to consider up to dimension-three operators in $J_E$. We find the following list of operators
\bea
j_{1,1}^{(2)} & =&-v^{2}\partial_{x}\theta\partial_{x}\phi\nonumber \\
j_{2,3}^{(2)} & =&\frac{i\varepsilon}{m}\left(d^{\dagger}\partial_{x}d+\bar d^{\dagger}\partial_{x}\bar d\right)\nonumber \\
j_{1,4}^{(2)} & =&-\frac{\alpha v^{2}}{\sqrt{\pi K}}\partial_{x}\theta\left(d^{\dagger}d-\bar d^{\dagger}\bar d\right)\nonumber \\
j_{3,4}^{(3)} & =&\frac{i\alpha v}{2m\sqrt{\pi K}}\partial_{x}\phi\left(d^{\dagger}\partial_{x}d-\partial_{x}d^{\dagger}d-\bar d^{\dagger}\partial_{x}\bar d+\partial_{x}\bar d^{\dagger}\bar d\right)\nonumber \\
j_{3,5}^{(3)} & =&-\frac{iV}{2m}\left[\bar d^{\dagger}\bar d\left(d^{\dagger}\partial_{x}d-\partial_{x}d^{\dagger}d\right)+d^{\dagger}d\left(\bar d^{\dagger}\partial_{x}\bar d-\partial_{x}\bar d^{\dagger}\bar d\right)\right]\nonumber \\
j_{1,6}^{(3)} & =&-4gv\partial_{x}\theta\partial_{x}\phi\left(d^{\dagger}d+\bar d^{\dagger}\bar d\right)\nonumber \\
j_{1,8}^{(3)} & =&-\mu_{+}v\partial_{x}\phi\partial_{x}\left(d^{\dagger}d-\bar d^{\dagger}\bar d\right)\nonumber \\
j_{1,9}^{(3)} & =&i\mu_{-}v\partial_{x}\phi\left(d^{\dagger}\partial_{x}d-\partial_{x}d^{\dagger}d-\bar d^{\dagger}\partial_{x}\bar d+\partial_{x}\bar d^{\dagger}\bar d\right)\label{eq:all_js}\\
j_{2,9}^{(2)} & =&2\mu_{-}\varepsilon\partial_{x}\theta\left(d^{\dagger}d-\bar d^{\dagger}\bar d\right)\nonumber \\
j_{4,9}^{(3)} & =&\frac{2\mu_{-}\alpha v}{\sqrt{\pi v}}\partial_{x}\theta\partial_{x}\phi\left(d^{\dagger}d+\bar d^{\dagger}\bar d\right)-\frac{i\mu_{-}\alpha v}{\sqrt{\pi v}}\left[d^{\dagger}d\left(\bar d^{\dagger}\partial_{x}\bar d-\partial_{x}\bar d^{\dagger}\bar d\right)+(d\leftrightarrow\bar d)\right]\nonumber \\
j_{2,11}^{(3)} & =&2i\vartheta_{-}\varepsilon\left[d^{\dagger}d\left(\bar d^{\dagger}\partial_{x}\bar d-\partial_{x}\bar d^{\dagger}\bar d\right)+(d\leftrightarrow\bar d)\right]\nonumber \\
j_{2,12}^{(3)} & =&2\kappa_{1}\varepsilon\partial_{x}\phi\partial_{x}\left(d^{\dagger}d-\bar d^{\dagger}\bar d\right)\nonumber \\
j_{2,13}^{(3)} & =&-2\kappa_{2}\varepsilon\partial_{x}\theta\partial_{x}\phi\left(d^{\dagger}d+\bar d^{\dagger}\bar d\right)\nonumber \\
j_{2,14}^{(3)} & =&-i\kappa_{3}\varepsilon\partial_{x}\phi\left(d^{\dagger}\partial_{x}d-\partial_{x}d^{\dagger}d-\bar d^{\dagger}\partial_{x}\bar d+\partial_{x}\bar d^{\dagger}\bar d\right).\nonumber
\eea

The calculation of  the commutators in Eq. (\ref{commJH}) is tedious but straightforward. To simplify the result, we use the known relations for the XXZ model $\varepsilon=1/m=v$. We find that the conservation law $[J_E,H]=0$ imposes the constraints    \begin{eqnarray}
g& = & -\frac{V}{4},\label{eq:vinc1}\\
g^\prime & = & 0,\\
\mu_{-} & = & -\frac{\alpha v}{2\sqrt{\pi K}}\label{eq:vinc2},\\
\mu_{+} & = & 2\kappa_{1}\label{eq:vinc3},\\
\kappa_{2} & = & -\frac{\alpha^{2}v}{2\pi K}\label{eq:vinc4},\\
\kappa_{3} & = & 0.
\end{eqnarray}
Most importantly, integrabiliy rules out the $g^\prime $ interaction. This is precisely the operator considered in \cite{castroneto} which accounts for a finite decay rate of a mobile impurity in a Luttinger liquid at finite temperatures.

\subsection{2. Excitation energies from  the thermodynamic Bethe ansatz}

In this section we describe the calculation of dressed energies using the thermodynamic Bethe ansatz (TBA) approach to the XXZ model \cite{takahashi72,takahashi}. 

Bethe ansatz states are parametrized by a set of rapidities $\{x_j^{\alpha}\}$ which satisfy the Bethe equations. Here $j$ labels the   type of  string and $\alpha$ specifies a particular rapidity.  More precisely, $x_j^\alpha$  refers to the  real part of the rapidity, since the imaginary part  is fixed by the string hypothesis \cite{takahashi72,takahashi}. Strings with length $n_j>1$ are interpreted as bound states of $n_j$ particles. For simplicity, we choose the anisotropy parameter to be $\Delta=\cos(\pi/\nu)$, with $\nu\in\mathbb{Z}$. In this case, we can restrict ourselves to a finite number of strings $j=1,2,\dots,\nu$. The strings with $j=1,2,\dots,\nu-1$ have length $n_j=j$ and parity $\upsilon_j=+1$; the string with $j=\nu$ has length $n_\nu=1$ and parity $\upsilon_\nu=-1$.  

The Bethe   equations (in logarithmic form) for a chain of length $N$ and periodic boundary conditions read\be
Nt_j(x_\alpha^j)=2\pi I_\alpha^j+\sum_{k=1}^{\nu}\sum_{\beta=1}^{M_k}\Theta_{jk}(x_\alpha^j-x_\beta^k),\qquad \alpha=1,\dots,M_j,\label{BAeq}
\ee
where $M_j$ is the number of strings of type $j$ in the Bethe ansatz state and $I_{\alpha}^j$ are integers (for $M_j$ odd) or half-integers (for $M_j$ even). A particular Bethe ansatz wave function is determined by the set of $I_\alpha^j$'s. The functions $t_j(x)$ and the scattering phase shifts  $\Theta_{jk}(x)$ are given by\bea
t_j(x)&=&f(x;n_j,v_j),\\
\Theta_{jk}(x)&=&f(x;|n_j-n_k|,v_jv_k)+f(x;n_j+n_k,v_jv_k)+2\sum_{l=1}^{ \textrm{Min}(n_j,n_k)-1}f(x;|n_j-n_k|+2l,v_jv_k),
\eea
where we define the function\be
f(x;n,v)=\left\{\begin{array}{cc}
0\,,& \textrm{ if } n/\nu\in \mathbb{Z}, \\
2v\arctan\left[(\cot(n\pi/2\nu))^v\tanh(\pi x/2\nu)\right],&\textrm{otherwise}.
\end{array}\right.
\ee

In the TBA approach, we take the   limit $N\to\infty$  and characterize the macroscopic state by the density of particles $\rho(x)$ and density of holes  $\rho^h(x)$ in rapidity space. The equilibrium state is obtained by minimizing the free energy as a functional of $\rho(x)$ and $\rho^h(x)$. This leads to a set of coupled nonlinear integral equations for the  dressed energies\be
\varepsilon_j(x)=-2\nu\sin(\pi/\nu)a_j(x)+\frac1{\beta}\sum_{k=1}^{\nu}\upsilon_k \int_{-\infty}^{+\infty}dy\,T_{jk}(x-y)\ln[1+e^{-\beta\varepsilon_k(y)}],\label{dressedenergies}
\ee
where $\beta=1/T$ is the inverse temperature and \bea
a_j(x)&=&\frac{1}{2\pi}\frac{dt_j}{dx},\\
T_{jk}(x)&=&\frac{1}{2\pi}\frac{d\Theta_{jk}}{dx}.
\eea
Eq. (\ref{dressedenergies}) can be solved numerically by iteration. In the limit $T\to 0$, the dressed energy for   $j=1$ (the even-parity one-string) reduces to the dispersion of the single-hole excitation over the ground state.

The dressed energies can be used to calculate the free energy and other thermodynamic properties \cite{takahashi}. They also show up as the energies of elementary excitations over the equilibrium state \cite{korepin}. The thermal excitation spectrum for the gapless phase of the XXZ model was calculated by Puga \cite{puga}. The  energy required to create a single hole  with rapidity $x$ in the density of type-$j$ strings is \be
\Delta E_j(x)=-\varepsilon_j(x)+\frac{1}{\beta}\sum_{k=1}^\nu \frac{\Theta_{kj}(\infty)}{\pi}\ln[1+e^{-\beta\varepsilon_k(\infty)}].
\ee 
Notice that the excitation energies $\Delta E_j(x)$ differ from the dressed energies in Eq. (\ref{dressedenergies}) by a constant term that involves all strings.  

\begin{figure}
\begin{center}
\includegraphics*[width=.35\columnwidth]{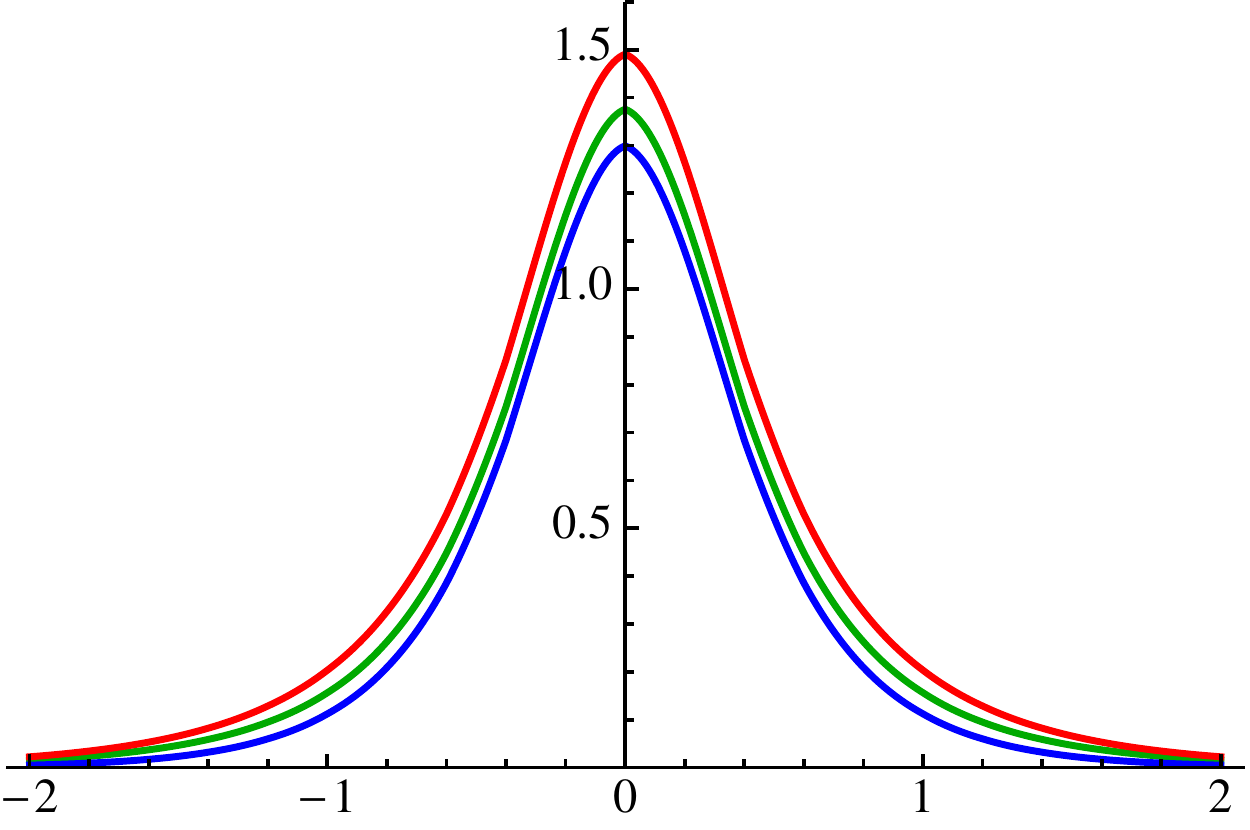}
\end{center}
\caption{Energy of single one-string excitation  as a function of rapidity $x$ for $\Delta=0.5$ and three different values of  temperature. From bottom to top: $T = 10^{-3}$, $T=0.25$, and $T= 5$. \label{fig:dressed}}
\end{figure}

Fig. \ref{fig:dressed} shows the excitation energy for the $j=1$
string for three different values of temperature. We are particularly
interested in the bandwidth, which is given by $\Delta E_1(0)$. At
$T=0$, the bandwidth is known analytically, $\lim_{T\to0}\Delta
E_1(0)=\frac{\pi\sqrt{1-\Delta^2}}{2\arccos\Delta}$. The important
point is that the bandwidth calculated from the TBA dressed energies
increases with temperature, contrary to the behavior of the
frequencies predicted by the effective field theory and observed
numerically using a tDMRG algorithm.

\subsection{3. Fits of the DMRG data}
We have fitted the tDMRG data using the fit function
\begin{equation}
\sqrt{t}G(t)=\Gamma+ A\left(\frac{\pi T}{\sinh(\pi T t)}\right)^{\tilde\eta}\text{e}^{-i(\tilde\varepsilon t+\tilde\varphi)}+B\, t^{-3/2}\text{e}^{-2i\tilde\varepsilon t}\text{e}^{-\rho t}
\end{equation}
where $A,B,\tilde\eta,\tilde\varepsilon,\tilde\varphi,\rho,\Gamma$ are
real fitting parameters. The first, constant term $\Gamma$ is the
diffusive contribution.  The second term is the single impurity
contribution, Eq.~(9) in the main text, while the third term
represents the two-impurity contribution, Eq.~(10), where we have
allowed for a small decay rate which seems to be of order
$\sim\text{e}^{-1/T}$ and might possibly be related to thermal
excitations at the band edges.  In table \ref{tab_suppl1} we present
the fit parameters for the fits shown in Fig.~2 and Fig.~3 of the main
text.
 \begin{table*}[!ht] 
 \begin{ruledtabular}
 \begin{tabular}{c|c|c|c|c|c|c|c|c|c|c}
 & $\Delta$& $T$ & $t$ fit range & $\Gamma$ & $A$ & $\tilde \eta$ & $\tilde\varepsilon$& $\tilde\varphi$ & $B$ & $\rho$\\\hline
 theory & 0.3 & 0 & --- & 0 & --- & 0.838 & 1.1835 & 0.530 & --- & ---\\
 theory, $\mathcal{O}(\Delta)$ & 0.3 & 0.07 & --- & $2.11\cdot 10^{-5}$ & --- & 0.831 & 1.1825 & 0.520 & --- & ---\\
 fit & 0.3 & 0.07 & $t\geq 10$ & $\sim 0$ & 0.275 & 0.796 & 1.185 & 0.751 & -0.202 & 0 \\
 theory, $\mathcal{O}(\Delta)$ & 0.3 & 0.25 & --- & $7.94\cdot 10^{-4}$ & --- & 0.814 & 1.1709 & 0.493 & --- & ---\\
 fit & 0.3 & 0.25 & $t\geq 10$ & $6.3\cdot 10^{-4}$ & 0.421 & 0.713 & 1.165 & 1.990 & -0.242 & 0.016\\\hline
 theory & 0.8 & 0 & --- & 0 & --- & 0.629 & 1.465 & 0.202 & --- & ---\\
 theory, $\mathcal{O}(\Delta)$ & 0.8 & 0.167 & --- & $5.58\cdot 10^{-4}$ & --- & 0.533 & 1.457 & 0.052 & --- & ---\\
 fit & 0.8 & 0.167 & $t\geq 15$ & $3.26\cdot 10^{-4}$ & 0.161 & 0.403 & 1.503 & -0.238 & 0 & --- \\
 theory, $\mathcal{O}(\Delta)$ & 0.8 & 0.25 & --- & $1.26\cdot 10^{-3}$ & --- & 0.492 & 1.448 & -0.013 & --- & ---\\
 fit & 0.8 & 0.25 & $t\geq 15$ & $9.88\cdot 10^{-4}$ & 0.210 & 0.387 & 1.525 & -0.397 & 0 & ---
 \end{tabular}
 \caption{\label{tab_suppl1} Parameters obtained by fitting the tDMRG data presented in the main text, see Figs.~2, 3.}
 \end{ruledtabular}
 \end{table*}

\begin{figure}
\begin{center}
\includegraphics*[width=.5\columnwidth]{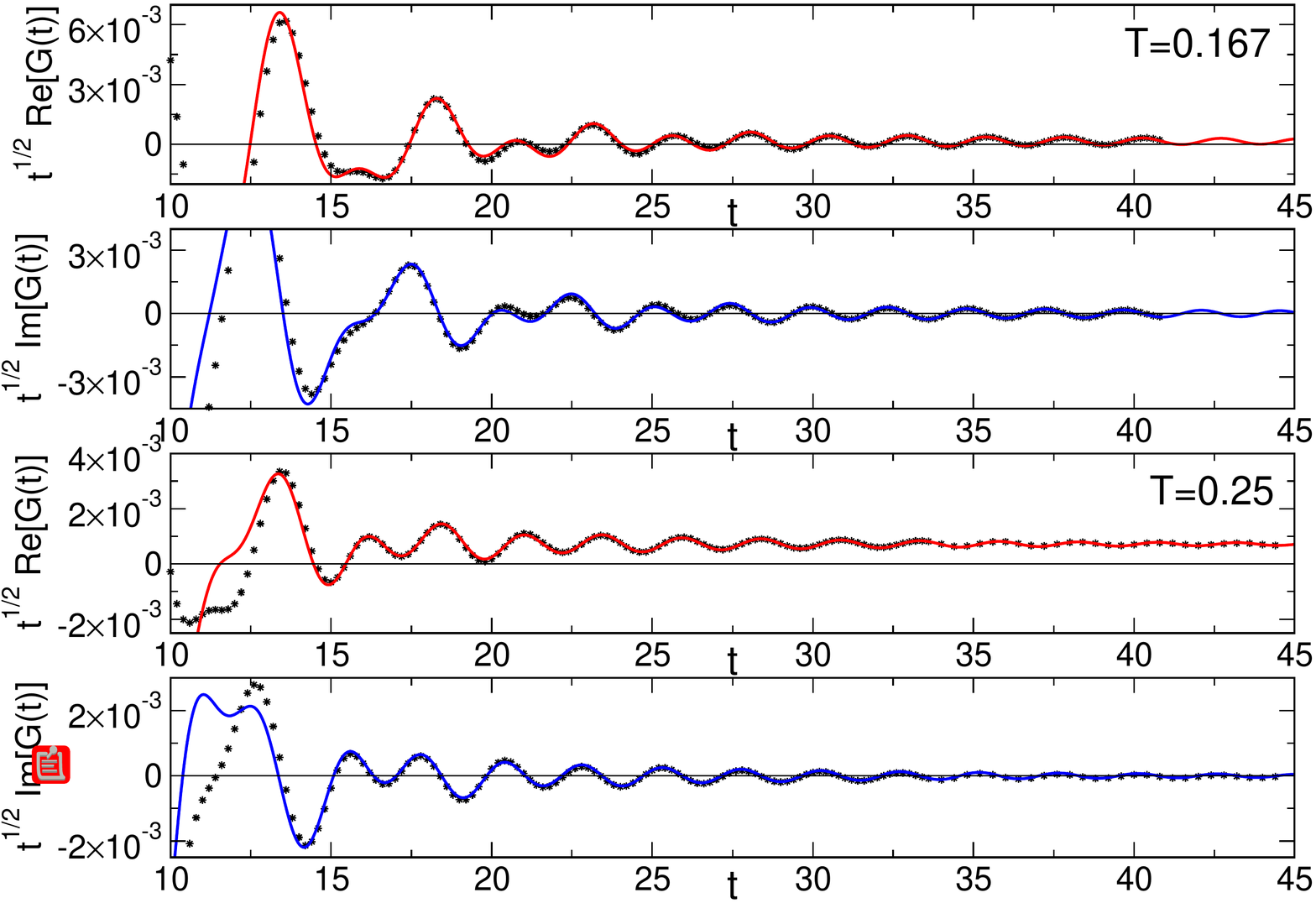}
\end{center}
\caption{$\sqrt{t}G(t)$ at $\Delta=0.5$: tDMRG data (symbols) and fits
  (lines) for $t\geq 15$. The fit parameters are given in table
  \ref{tab_suppl2}.}
\label{Fig_suppl2}
\end{figure}

 One of our main findings based on the analysis of the numerical data
 is that the two-impurity contribution becomes very small for large
 interaction strengths leading to a fit parameter $B$ for $\Delta=0.8$
 which is essentially zero. To further support that $B$ for the
 considered temperatures is strongly reduced with interaction, we
 present in Fig.~\ref{Fig_suppl2} below tDMRG data and fits for
 intermediate interaction strength $\Delta=0.5$.
Here a two-impurity contribution is still visible but the amplitude
$B$ is already very small, see table \ref{tab_suppl2}. In order to
illustrate the sensitivity of the fit parameters on the fit interval
we concentrate on the case $\Delta=0.5$, $T=0.167$ and show in table
\ref{tab_suppl2} parameters for fits using three different time
intervals.
\begin{table*}[!ht] 
\begin{ruledtabular}
\begin{tabular}{c|c|c|c|c|c|c|c|c|c|c}
& $\Delta$& $T$ & $t$ fit range & $\Gamma$ & $A$ & $\tilde \eta$ & $\tilde\varepsilon$& $\tilde\varphi$ & $B$ & $\rho$\\\hline
theory & 0.5 & 0 & --- & 0 & --- & 3/4 & 1.30 & 0.393 & --- & ---\\
theory, $\mathcal{O}(\Delta)$ & 0.5 & 0.167 & --- & $3.98\cdot 10^{-4}$ & --- & 0.710 & 1.2919 & 0.329 & --- & ---\\
fit 1 & 0.5 & 0.167 & $t\geq 10$ & $1.68\cdot 10^{-4}$ & 0.277 & 0.575 & 1.294 & 0.853 & -0.086 & 0.019 \\
fit 2  & 0.5 & 0.167 & $t\geq 15$ & $1.49\cdot 10^{-4}$ & 0.199 & 0.527 & 1.287 & 1.406 & -0.099 & 0.020 \\
fit 3  & 0.5 & 0.167 & $t\geq 20$ & $1.51\cdot 10^{-4}$ & 0.284 & 0.558 & 1.285 & 1.853 & -0.093 & 0.019 \\
theory, $\mathcal{O}(\Delta)$ & 0.5 & 0.25 & --- & $1.09\cdot 10^{-3}$ & --- & 0.690 & 1.2829 & 0.299 & --- & ---\\
fit   & 0.5 & 0.25 & $t\geq 15$ & $7.19\cdot 10^{-4}$ & 0.266 & 0.501 & 1.272 & 2.196 & -0.091 & 0.041 
\end{tabular}
\caption{\label{tab_suppl2} Parameters for various fits of the tDMRG data at $\Delta=0.5$.}
\end{ruledtabular}
\end{table*}
Except for the phase shift $\tilde\varphi$, and, to a lesser extent,
the amplitude $\tilde A$, all fit parameters show little variation
implying, in particular, that it is possible to extract the
temperature dependence of $\tilde\varepsilon(T)$ and $\tilde\eta(T)$
with reasonably accuracy from the tDMRG data. On the other hand, we
want to emphasize that the phase shift $\tilde\varphi$ cannot be fixed
reliably from numerical data even at zero temperature, see
Refs.~\cite{pereira2008,pereira2009}.

\end{document}